\documentstyle[12pt]{article}

\newcommand{\Dslash}{\not \kern-4pt D}
\newcommand{\dslash}{\not D}
\newcommand{\Aslash}{\not \kern-5pt A}
\newcommand{\Eslash}{\not \kern-5pt \partial}
\newcommand{\Tslash}{- \kern-4 T}

\begin{document}

\title{Fermionic Determinant of the Massive Schwinger Model}

\author{M.P.Fry\\[2mm]
School of Mathematics\\
Trinity College\\
Dublin 2\\
Ireland}

\date{}

\maketitle

\vfill

\begin{abstract}
A representation for the fermionic determinant of the massive Schwinger
model, or $QED_2$, is obtained that makes a clean separation between the
Schwinger model and its massive counterpart. From this it is shown that
the index theorem for $QED_2$ follows from gauge invariance, that the
Schwinger model's contribution to the determinant is canceled in the
weak field limit, and that the determinant vanishes when the field
strength is sufficiently strong to form a zero-energy bound state.
\end{abstract}

\vfill

\newpage

Quantum electrodynamics in two dimensional space time
($QED_2$),\,otherwise
known as the massive Schwinger model,\,is defined in Euclidean space
by the action
\begin{equation}
 S[A,\bar{\Psi},\Psi]=\frac{1}{2}\int
  d^{2}xB^2+\int d^2x\bar{\Psi}(\Dslash+m)\Psi,
\end{equation}
where $\Dslash=\vec{\gamma}.(-i\vec{\bigtriangledown}-e\vec{A})$ and
$B=F_{01}=\partial_0A_1-\partial_1A_0$.  Our designation of $F_{01}$ as
a magnetic field is consistent with regarding $S$ as the Hamiltonian for
a charged,\,massive fermion confined to a plane in the presence of a
static magnetic field perpendicular to the plane.  For definiteness we
set $\gamma_0=-i\sigma_1,\gamma_1=-i\sigma_2$,\,where $\sigma_{1,2}$ are
the Pauli matrices.The model is super-renormalizable,\,requiring no
infinite renormalization other than a trivial renormalization of the
zero-point energy.  Hence $e$ and $m$ are finite parameters.

The case when $m$=0,\,known as the Schwinger model [1],\,is exactly
soluble and has become an important tool for gaining insight into gauge
field theories.\,It continues to generate enormous interest with some
fifty preprints per annum connected to the model and variations of
it.\,The literature for the case $m\not=0$ is sparse,\,the classic
references remaining those in [2].\,It is not thought to be exactly
soluble.\,As might be suspected by our interpretation of the massive
model's action,\,its fermionic determinant determines\,(after
integrating over the fermion mass) the one-loop effective action for
$QED_4$ in the presence of smooth,
\,polynomial-bounded,\,unidirectional,\,static magnetic fields with fast
decrease at infinity [3]. Therefore, $QED_2$ contains information on
physics in four dimensions and should not be regarded as just a model.

In this Letter we wish to consider $QED_2$'s gauge invariant fermionic
determinant. It will appear in the computation of the theory's n-point
functions as a result of integration over the fermionic degrees of
freedom. The first problem is to make sense out of the formal
expressions
\begin{equation}
{\det}^2(1-S\Aslash)
  =\frac{\det[(\vec{p}-\vec{A})^2-\sigma_3B+m^2]}{\det[p^2+m^2]},
\end{equation}
on a Euclidean manifold. Here S is the free (Euclidean)
fermion-propagator, and $e$ has been absorbed into $A_{\mu}$. There are
several ways to define determinants of Dirac operators [4], but one of
these definitions seems more suited than others to grasp the known
simplifications presented by $QED_2$, namely, the ''propertime
regularization`` definition [5]. It defines the determinant as
\begin{equation}
 \ln\det(\Dslash^{\dagger}\Dslash+m^2)
  =-\int^{\infty}_{\epsilon}
  \frac{ds}{s}Tr[\exp(-s\Dslash^\dagger\Dslash)]e^{-sm^2},
\end{equation}
where $\epsilon$ is an ultraviolet cutoff which, due to
super-renormalizability, can be set to zero later. Because we will
always assume $m^2>0$, we feel assured that potential infrared
divergences due to the zero modes of $\Dslash^{\dagger}\Dslash$ are
regulated.

The above definition of the determinant respects gauge invariance.
Therefore we should be able to calculate in the Lorentz gauge
$\partial_\mu A_\mu=0$ which, in two dimensions, allows us to set
$A_\mu=\epsilon_{\mu\nu}\partial_{\nu}\phi$, with $B=-\partial^2\phi$
and $\Aslash=i\sigma_3\Eslash\phi$. The antisymmetric tensor
$\epsilon_{\mu\nu}$ is normalized as $\epsilon_{01}=1$. Following
Alvarez [5], we consider the operator
\begin{equation}
\Dslash_t=-i\Eslash-t\Aslash
  =-ie^{-t\sigma_3\phi}\Eslash e^{-t\sigma_3\phi},
\end{equation}
where t is a real parameter. Differentiating with respect to t,
\begin{equation}
\dot{\Dslash}=-\sigma_3\phi\Dslash_t-\Dslash_t\sigma_3\phi,
\end{equation}
we calculate
\begin{eqnarray}
\frac{d}{dt}\ln\det(\Dslash_t^{\dagger}\Dslash_{t}+m^2)
 & = & 4\int^\infty_\epsilon\!ds
       Tr(\sigma_3\phi\Dslash^2_te^{s\dslash^2_t})
       e^{-sm^2} \nonumber \\
 & = & -4Tr(\sigma_3\phi e^{\epsilon\dslash^2_t})
       e^{-\epsilon m^2} \nonumber \\
 &   & +4m^2\int^\infty_\epsilon\!ds
       Tr(\sigma_3\phi e^{s\dslash^2_t})e^{-sm^2}.
\end{eqnarray}
Noting that for small $\epsilon$,
\begin{equation}
<x|e^{\epsilon\dslash_{t}^2}|x>
  =\frac{1}{4\pi\epsilon}
  (1-\epsilon t\sigma_3\partial^2\phi + O(\epsilon^2)),
\end{equation}
we obtain our definition of the fermionic determinant,
\begin{eqnarray}
\ln\left[\frac{\det(\Dslash^\dagger\Dslash + m^2)}
              {\det(p^2 + m^2)}\right]^\frac{1}{2}
&= & \frac{1}{2\pi}\int\!d^2x\phi \partial^2 \phi
  \nonumber \\
& + & 2m^2\int^{1}_{0}\,dtTr\{[(H_{+}^{(t)}+m^2)^{-1}
   - (H^{(t)}_{-}+m^2)^{-1}]\phi\},
 \nonumber \\
& &
\end{eqnarray}
where $H_{\pm}^{(t)}=(\vec{P}-t\vec{A})^2{\mp}tB$. Note that this
definition makes a clean separation between the Schwinger model, the
first term, and its massive counterpart. Its perturbrative expansion in
powers of e is consistent with known results. Thus, it reproduces the
$O(e^2)$ result for the vacuum polarization graph
\begin{eqnarray}
\ln \det & = & \frac{1}{2\pi}\int\phi\partial^2\phi + 2m^2\int\!d^2x<x|\phi
               \frac{1}{p^2+m^2}B\frac{1}{p^2+m^2}|x> \nonumber \\
         & = & -\frac{1}{2\pi}
               \int\frac{d^2q}{(2\pi)^2}|\hat{B}(q)|^2\int^{1}_{0}
               \!dz\frac{z(1-z)}{q^2z(1-z)+m^2},
\end{eqnarray}
where $\hat{B}$ is the Fourier transform of $B$. In addition, graphs of
$O(e^{4})$ and higher vanish order by order in the limit $m^{2}=0$, in
accordance with Schwinger's original result [1].

We have not integrated by parts in the first term of Eq.(8) as is
usually done. In the Lorentz gauge the auxiliary potential
$\phi(\vec{x})=-\int\!d^2yln|\vec{x}-\vec{y}|B(\vec{y}) / 2\pi$ and,
assuming that the flux $\Phi=\int\!d^2xB\not = 0$, integration by parts
is not justified here.

It is by now evident that we are assuming our potentials are
sufficiently smooth with enough fall off at infinity so that everything
we have done makes mathematical sense. But note: if $\Phi\not=0$, $A_\mu$
in the Lorentz gauge behaves like a ``winding''field with a $1/|x|$ fall
off. This will have some consequences below. It might be objected that
since $A_\mu$ is to be integrated over, it should be a random field. Our
strategy is to first calculate the determinant in an external field in a
convenient gauge, the Lorentz gauge, assuming nice potentials, then
switching to whichever gauge and potentials are best suited for making
sense out of the remaining integration over $A_\mu$.  Of course, any
gauge-invariant constraints imposed on the determinant required, say, to
make it non-vanishing, have to be honored by the functional integral.

Within the Lorentz gauge we still have the freedom to shift $\phi$ by a
constant$:\phi \to \phi + c$. By definition, the determinant depends on
$A_\mu$ and so is invariant under this shift. Referring to Eq.(8), we
have consistency provided
\begin{equation}
e^2\Phi/2\pi=2m^2e\int^{1}_{0}\!dt
   Tr[(H_{+}^{(et)}+m^2)^{-1}-(H_{-}^{(et)}+m^2)^{-1}],
\end{equation}
where we have temporarily restored the coupling $e$. We can get rid of
the t-integration by setting $\lambda=et$ and differentiating both sides
with respect to e . The result is
\begin{equation}
 \Phi/2\pi=m^{2}Tr[(H_{+}+m^{2})^{-1}-(H_{-}+m^{2})^{-1}],
\end{equation}
where we have again absorbed $e$ into $A_\mu$ and $B$ and set
$H_{\pm}=(\vec{P}-\vec{A})^2 \mp B$. But the right-hand side of Eq.(11)
is independent of $m^2$ [6]. One way to see this is to rewrite the
right-hand side as
$$
m^2\int^{\infty}_{0}\!dse^{-sm^2}Tr(e^{-sH_{+}}-e^{-sH_{-}}),
$$
and appeal to the supersymmetry of the operator pair $H_{\pm}$ [7] so
that only the zero modes of $H_{\pm}$ contribute. This way regulating
the trace in Eq.(11) in fact follows from our definition of the
determinant [see last term in Eq.(6)] and serves as a reminder of how to
deal with any doubt about the trace operation. Thus, gauge invariance
leads to the condition

\begin{eqnarray}
\Phi/2\pi & = & Tr[P_{+}(0)-P_{-}(0)] \nonumber \\
  {  }    & = & n_{+}-n_{-}+\frac{1}{\pi}\sum_{l}[\delta_{+}^{l}(0)-
                        \delta_{-}^{l}(0)],
\end{eqnarray}
where $P_{\pm}(0)$ are projection operators into the subspace of
zero-energy modes of $H_{\pm}$; $n_{\pm}$ denote the number of
zero-energy bound states of $H_{\pm}$, and $\delta^{l}_{\pm}(0)$ are the
zero-energy phase shifts for scattering by the Hamiltonians $H_{\pm}$ in
a suitable angular momentum basis l.  Equation (12) is just the index
theorem for a two-dimensional Euclidean manifold [8,9]. By the
Aharonov-Casher theorem [7,10] we know that $n_{+}(n_{-})$ is
$\{|\Phi|/2\pi\}$, all with $\sigma_{3}=1 (\sigma_{3}=-1)$ if $\Phi>0
(\Phi<0)$. Here $\{x\}$ denotes the largest integer strictly less than
$x$ and $\{0\}=0$. This is our first result, that the index theorem for
$QED_2$ follows from gauge invariance.

Let us now write Eq.(8) in the form
\begin{eqnarray}
\ln\det &=& -\frac{1}{2\pi}
            \int\!\!d^2x\phi B+2\int^{1}_{0}
            \!dtTr\{[P_{+}^{(t)}(0)-P_{-}^{(t)}(0)]\phi\}
        \nonumber \\
        && +2m^2\int^{1}_{0}\!dtTr'\{[H_{+}^{(t)}+m^2)^{-1}
           -(H_{-}^{(t)}+m^2)^{-1}]\phi\},
\end{eqnarray}
where $P^{(t)}_{\pm}(0)$ are projection operators into the subspace of
zero-energy modes of $H^{(t)}_{\pm}$. The prime on the second trace
symbol indicates that zero modes are omitted.  Now consider magnetic
fields such that $|\Phi|/2\pi\leq 1$ so that there are no bound states.
According to Musto et al. [9] we can write the second term in Eq.(13) as
$$
\frac{2}{\pi}\int^{1}_{0} dt
  Tr\{ [\delta_{+}^{(t)}(0)-\delta_{-}^{(t)}(0)]\phi \},
$$
where the trace is over scattering states, in the limit of zero energy,
of the free Hamiltonian $H_{0}$ defined by
$H^{(t)}_{\pm}=H_{0}+V^{(t)}_{\pm}$. The operators $\delta_{\pm}^{(t)}$
are calculated from the S-matrix $S(\lambda)=\exp(2i\delta(\lambda))$ as
$\lambda \downarrow 0$. Let us assume further that the magnetic field is
sufficiently weak to justify the first Born approximation
\begin{eqnarray}
\delta_{+}^{(t)}(0)-\delta_{-}^{(t)}(0)
 & = & -\pi\delta(H_{0})(V_{+}^{(t)}-
       V_{-}^{(t)})  \nonumber \\
 & = & 2\pi tB\delta(H_0).
\end{eqnarray}
The normalized eigenfunctions of $H_0$ are
$\psi_{El}{\vec{(r)}}
           = J_{l}(kr)e^{il\theta}/\sqrt{4\pi}$.  Then
\begin{eqnarray}
\frac{2}{\pi}\int^{1}_{0}\!dt
     Tr\{[\delta_{+}^{(t)}(0)-\delta_{-}^{(t)}(0)]\phi\}
           & = & 4\int^{1}_{0}\!dtt\int^{\infty}_{0-}\!dE
            \sum^{\infty}_{l=-\infty}<El|\delta(H_0)B\phi |El>
                  \nonumber \\
      & = & \frac{1}{2\pi}\int^{\infty}_{0-}\!dE\delta(E)\int\!d^{2}r
        \sum^{\infty}_{l=-\infty}J^{2}_{l}(kr)B(\vec{r})\phi(\vec{r})
                \nonumber \\
           & = & \frac{1}{2\pi}\int\!d^{2}xB\phi,
\end{eqnarray}
where we used the identity $\sum^{\infty}_{l=-\infty}J^{2}_{l}(x)=1$.
This result cancels the first term in Eq.(13) and is our second result,
namely that the Schwinger model's contribution to the determinant of
$QED_2$ is canceled in first Born approximation by a contribution from
the zero modes in the massive sector.  It may be that our weak-field
approximation to the second term in Eq.(13) is exact for
$|\Phi|/2\pi \leq 1$, but we have not been able to prove this.

Finally, let us increase the magnetic field to $|\Phi|/2\pi>1$ so that
zero-energy bound states of $H_{\pm}^{(t)}$ begin to appear. These
states are of the form [7,11] $\psi^{(t)}(x,y)=f_{\pm}\exp(\pm t \phi)$,
where $f_{\pm}$ are t-independant polynomials in $x\pm iy$ of degree
$< |\Phi |t/2\pi -1 $, and $\phi$ is the auxiliary potential defined
above. These zero modes are not in general orthonormal. We define the
norm matrix $N_{ij}(t)=(\psi_{i}^{(t)},\psi_{j}^{(t)})$ and the
projection kernel on the zero-mode $L^2$ subspace~[12]
\begin{equation}
P^{(t)}(x,y)=\sum^{n}_{i,j=1}
  \psi_{i}^{(t)}(x)(N^{-1}(t))_{ij}(\psi_{j}^{(t)})^{\dagger}(y),
\end{equation}
with $TrP^{(t)}=n\equiv\{|\Phi|t/2\pi\}$. As previously noted, the bound
states all have the same chirality, depending on the sign of $\Phi$.
Their contribution to the second term in Eq.(13) is, for
$|\Phi|/2\pi>1$,
\begin{eqnarray}
&& \hspace{-31 mm}
  \pm2\int^{1}_{0}\!dt
   \sum^{n}_{i,j=1}(N^{-1}(t))_{ij}
    \int\!d^{2}xf_{i}f_{j}^{*}\phi e^{\pm2t\phi}  \nonumber \\
& = & \pm2\lim_{\epsilon \downarrow 0}
        [\int^{4\pi/|\Phi|}_{2\pi(1+\epsilon)/|\Phi|}\;dt
        \; N^{-1}(t)\int\!d^{2}x\phi e^{\pm2t\phi} \nonumber \\
&   &  +\int^{6\pi/|\Phi|}_{4\pi(1+\epsilon)/|\Phi|}\;dt
        \; \sum^{2}_{i,j=1} (N^{-1}(t))_{ij}
        \int\!d^{2}xf_{i}f_{j}^{*}\phi e^{\pm2t\phi}+\cdots]
\end{eqnarray}
The above integrals can be expressed in terms of the norms $N_{ij}$ and
their derivatives with respect to $t$ after an integration by parts in
$t$.  The result is the following zero-energy bound state contribution
to the second term in Eq.(13):
\begin{eqnarray*}
\lim_{\epsilon \downarrow 0}\ln\biggl[
&&
\frac{N(\frac{4\pi}{|\Phi|})}{N(\frac{2\pi(1+\epsilon)}{|\Phi|})}
  \\
&& \times
          \frac{\det N_{ij}\left( \frac{ 6\pi}{|\Phi| }\right) }
            { \det N_{ij}\left(\frac{ 4\pi(1+\epsilon) }{ |\Phi|}\right)}
       \biggr|^{2}_{i,j=1}
   \\
&& \cdots
   \\
&& \times
 \frac{\det N_{ij}(1)}
  {\det N_{ij}\left(2\pi\left\{\frac{|\Phi|}{2\pi}\right\}
        (1+\epsilon)/{|\Phi|}\right)}
  \biggr|_{i,j=1}^{\{|\Phi|/2\pi\}} \biggr].
\end{eqnarray*}
The norm of the first bound state, occuring when $2\geq|\Phi|/2\pi>1$,
diverges as $\epsilon \downarrow 0$:
\begin{eqnarray}
N(2\pi(1+\epsilon)/|\Phi|) & = & \int\!d^{2}x\exp[\pm4\pi(1+\epsilon)
                                      {\phi/|\Phi|}] \nonumber\\
                            & = & 2\pi\int^{\infty}_{R}\!drr\exp[-2
                                      (1+\epsilon)\ln r]+\mbox{finite at }
                                      \epsilon=0, \nonumber\\
& &
\end{eqnarray}
where $R$ is large compared to the range of $B$. Hence the logarithm in
the expression displayed between Eqs. (17) and (18) becomes minus
infinity, thereby causing a zero to appear in the fermionic determinant
of $QED_2$, as seen from Eq.(13). This is our third result. Including
more bound states does not improve matters. The problem seems to lie
with the slow ${1/|x|}$ fall off of $A_{\mu}$ in the Lorentz gauge when
$\Phi \not = 0$.

We have always kept $m^2>0$. If we take the limit $m^2=0$, it appears
from the foregoing that the zero-mass limit of the fermionic determinant
of $QED_2$ does not converge uniformly to that of the Schwinger model,
which was calculated with $m^2=0$ ab initio. This statement is made
subject to the proviso that the $m^2=0$ limit is taken before an
expansion in powers of e is made; otherwise, as previously noted, we do
indeed regain the Schwinger model's determinant if we take the $m^2=0$
limit order by order.

More questions have been raised here than answered, but our results do
indicate that $QED_2$ remains a rich and relatively unexplored source of
physics.

The author wishes to thank L.O'Raifeartaigh, S.Sen and J.Sexton for useful
discussions.

\end{document}